\documentclass[amsmath,amssymb,aps,prl,twocolumn,showkeys,superscriptaddress,10pt,floatfix,showpacs]{revtex4-1}

\usepackage{tabularx}
\usepackage{makecell}
\usepackage[dvipsnames]{xcolor}
\usepackage{fdsymbol}
\usepackage{graphicx}
\usepackage{bm}
\usepackage{siunitx}
\usepackage{hyperref}

\usepackage{xcolor}
\hypersetup{
    colorlinks,
    linkcolor={red!50!black},
    citecolor={blue!50!black},
    urlcolor={blue!80!black}
}

\newcommand{\Cs}{CsEuFe$_4$As$_4$}
\newcommand{\Rb}{RbEuFe$_4$As$_4$}
\newcommand{\CaK}{CaKFe$_4$As$_4$}
\newcommand{\FeAs}{Fe$_4$As$_4$}
\newcommand{\Eu}{EuFe$_2$As$_2$}
\newcommand{\Tc}{$T_c$}
\newcommand{\Tm}{$T_m$}
\newcommand{\TN}{$T_N$}
\newcommand{\Pc}{$P_c$}
\newcommand{\acfirst}{$\chi^{\prime}$}
\newcommand{\acthird}{$\chi_3^{\prime}$}

\begin{document}
\title{\texorpdfstring{Superconducting and magnetic phase diagram of\\ RbEuFe$_4$As$_4$ and  CsEuFe$_4$As$_4$ at high pressure}{Superconducting and magnetic phase diagram of RbEuFe4As4 and  CsEuFe4As4 at high pressure}}
\author{Daniel\ E.\ Jackson}
\author{Derrick VanGennep}
\affiliation{Department of Physics, University of Florida, Gainesville, FL 32611, USA}
\author{Wenli Bi}
\affiliation{Advanced Photon Source, Argonne National Laboratory, Argonne, IL 60439, USA}
\affiliation{Department of Geology, University of Illinois at Urbana-Champaign, Urbana, IL 61801, USA}
\author{Dongzhou Zhang}
\affiliation{Hawaii Institute of Geophysics and Planetology, School of Ocean and Earth Science and Technology, University of Hawaii at Manoa, Honolulu, HI 96822, USA}
\author{Philipp Materne}
\affiliation{Advanced Photon Source, Argonne National Laboratory, Argonne, IL 60439, USA}
\author{Yi Liu}
\author{Guang-Han Cao}
\affiliation{Department of Physics, Zhejiang University, Hangzhou 310027, China}
\author{Samuel\ T.\ Weir}
\affiliation{Physics Division, Lawrence Livermore National Laboratory, Livermore, CA 94550, USA}
\author{Yogesh\ K.\ Vohra}
\affiliation{Department of Physics, University of Alabama at Birmingham, Birmingham, AL, 35294, USA}
\author{James\ J.\ Hamlin}
\email{Corresponding author: jhamlin@ufl.edu}
\affiliation{Department of Physics, University of Florida, Gainesville, FL 32611, USA}

\begin{abstract}
The recently discovered (Rb,Cs)EuFe$_4$As$_4$ compounds exhibit an unusual combination of superconductivity ($T_c \sim \SI{35}{K}$) and ferromagnetism ($T_m \sim \SI{15}{K}$).
We have performed a series of x-ray diffraction, ac magnetic susceptibility, dc magnetization, and electrical resistivity measurements on both \Rb\ and \Cs\ to pressures as high as $\sim \SI{30}{GPa}$.
We find that the superconductivity onset is suppressed monotonically by pressure while the magnetic transition is enhanced at initial rates of $dT_m/dP \sim \SI{1.7}{K/GPa}$ and $\SI{1.5}{K/GPa}$ for \Rb\ and \Cs\ respectively.
Near \SI{7}{GPa}, \Tc\ onset and \Tm\ become comparable.
At higher pressures, signatures of bulk superconductivity gradually disappear.
Room temperature x-ray diffraction measurements suggest the onset of a transition from tetragonal (T) to a half collapsed-tetragonal (hcT) phase at $\sim \SI{10}{GPa}$ (\Rb) and $\sim \SI{12}{GPa}$ (\Cs).
The ability to tune \Tc\ and \Tm\ into coincidence with relatively modest pressures highlights (Rb,Cs)EuFe$_4$As$_4$ compounds as ideal systems to study the interplay of superconductivity and ferromagnetism.
\end{abstract}

\maketitle

\section{Introduction}
The iron-based superconductors crystallize in several closely related crystal structures.
The most recently discovered of these is the so-called ``1144'' structure type, which was reported for compounds with the formula $AeA$Fe$_4$As$_4$, where $Ae$ = Ca or Sr and $A$ = K, Rb, or Cs~\cite{Iyo2016}.
These structures can be viewed as an ordered stacking of Fe$_2$As$_2$ layers sandwiched between alternating layers of $Ae$ and $A$.
Unlike the closely related ``122'' compounds such as BaFe$_2$As$_2$, which require doping or pressure to exhibit high \Tc\ values, the stoichiometric 1144 compounds exhibit $T_c \sim \SI{35}{K}$ at ambient pressure.
Subsequently, it was found that the same structure type forms when the alkaline earth element is replaced by the rare earth element Eu, resulting in \Rb\ and \Cs~\cite{Kawashima2016,Liu2016b,Liu2016a}.

The Eu variants of the 1144 structure exhibit an unusual coexistence of superconductivity ($T_c \sim \SI{35}{K}$) and what is nominally ferromagnetism ($T_m \sim \SI{15}{K}$).
The large ordered moment of $\sim$ 6.5 $\mu_B$ per formula unit is consistent with magnetism deriving from Eu$^{2+}$ ions~\cite{Liu2016b}.
M\"{o}ssbauer spectroscopy measurements confirm that the magnetism derives from Eu$^{2+}$ moments (which orient perpendicular to the crystallographic $c$-axis), and indicate that there is no magnetic order of Fe moments down to at least \SI{2}{K}~\cite{Albedah2018,Albedah2018b}.
A recent pre-print concluded that the Eu magnetism is quasi-2D in nature with strong magnetic fluctuation effects~\cite{Smylie2018}.
However, there are as yet no reports of \textit{e.g.,} neutron scattering measurements to establish the magnetic structure, so it is possible that the order is a more complicated modulated structure, rather than simple ferromagnetism~\cite{Sinha1982,Burlet1986}.

It is thought that superconductivity and magnetism are able to coexist in these compounds because of a weak coupling between the Eu planes and the FeAs planes.
Weak coupling between superconductivity and the Eu magnetism is confirmed by the fact that nonmagnetic CaRb\FeAs~\cite{Iyo2016} exhibits nearly the same \Tc\ as \Rb\ and \Cs.
Furthermore, a study of the (Eu$_{1-x}$Ca$_x$)Rb\FeAs\ series found that while \Tm\ vanishes with increasing Ca concentration, \Tc\ remains essentially constant~\cite{Kawashima2018}.
The observations described above place the (Rb,Cs)Eu\FeAs\ compounds in the class of local moment ferromagnetic superconductors such as ErRh$_4$B$_4$~\cite{Fertig1977}  and Ho$_x$Mo$_6$S$_8$~\cite{Ishikawa1977}.
However, these materials exhibit a destruction of the superconducting state at the onset of ferromagnetism, unlike in (Rb,Cs)Eu\FeAs.
In contrast, the uranium-based superconducting ferromagnets such as URhGe~\cite{Aoki2001}, UGe$_2$ (under pressure)~\cite{Saxena2000}, and UCoGe~\cite{Huy2007} show the onset of weak itinerant ferromagnetism at temperatures above the superconducting $T_c$.
We use the terminology ferromagnetic superconductor (FMS) to indicate $T_c > T_m$ and superconducting ferromagnet (SFM) to indicate $T_m > T_c$.

Experiments aimed at tuning the superconducting and magnetic transitions via pressure, chemical substitution, and applied magnetic fields have played a central role in our understanding of magnetic superconductors~\cite{Kakani1988,Wolowiec2015}.
To date, there appear to be only two reports of  chemical substitution experiments on (Rb,Cs)Eu\FeAs\ compounds.
In addition to the Eu $\rightarrow$ Ca, substitution study described above~\cite{Kawashima2018}, Liu \textit{et al.} explored the effect of Ni substitution on the Fe site~\cite{Liu2017}.
Increasing Ni concentration has little impact on the Eu magnetism, but produces the emergence of possible spin density wave order at 5\% Ni, a crossover from FMS to SFM near 6.5\% Ni, and finally the disappearance of superconductivity above 8\% Ni.
The rapid suppression of \Tc\ was attributed partly to the disorder induced by Ni substitution.

High pressure experiments have the potential to tune \Tc\ and \Tm\ without introducing intrinsic disorder.
In this paper we report a series of high pressure measurements on polycrystalline samples of both \Rb\ and \Cs .
Using a combination of x-ray diffraction, dc magnetization, ac magnetic susceptibility, and electrical resistivity measurements, we have mapped the phase diagrams of both compounds to pressures as high as $\sim \SI{30}{GPa}$

\section{Experimental Methods}
Polycrystalline samples of \Rb\ and \Cs\ were synthesized as previously reported~\cite{Liu2016b,Liu2016a}.
These samples were subjected to a variety of high pressure measurements, each using different diamond anvil cells (DAC).
Pressure was determined via the fluorescence spectrum of small pieces of ruby located inside the sample chamber, near the sample~\cite{Chijioke2005}.
For measurements at low temperature, the pressure was measured \textit{in situ} at low temperature, thus avoiding potential errors in pressure determination due to the changes in pressure that typically occur on cooling from room to low temperature.

Angle-dispersive x-ray diffraction (XRD) experiments on \Rb\ and \Cs\ powder samples were carried out at beamline 13BM-C (PX$^{\wedge}$2), Advanced Photon Source (APS), Argonne National Laboratory~\cite{Zhang2017}.
The X-ray beam was monochromated with silicon (311) to \SI{28.6}{keV} (\SI{0.434}{\angstrom}) with \SI{1}{eV} bandwidth.
A Kirkpatrick-Baez mirror system was used to focus the beam to $\SI{20}{\um} (\mathrm{vertical}) \times \SI{15}{\um} (\mathrm{horizontal})$ (FWHM) at the sample position.
The MAR165 Charge-Coupled Device (CCD) detector (Rayonix) was used to collect diffraction patterns.
Powdered LaB$_6$ was used to calibrate the distance and tilting of the detector. 

For the diffraction measurements, high pressure was achieved in Mao-type symmetric DACs with c-BN seats to allow access to high diffraction angles.
Two experimental runs were performed on \Cs.
In the first run, a pair of diamond anvils with \SI{600}{\um} culet were used up to \SI{11.1}{GPa}.
A Re gasket was pre-indented from \SI{250}{\um} initial thickness to \SI{85}{\um}.
During the second run, a pair of \SI{500}{\um} diamond anvils with Re gasket thickness of \SI{80}{\um} were used to achieve pressures up to \SI{28.1}{GPa}.
A single experimental run was performed on the \Rb\ sample, using diamond anvils of \SI{500}{\um} culet up to \SI{29.7}{GPa}.
The Re gasket was pre-indented to \SI{78}{\um} and a hole of \SI{260}{\um} was EDM drilled.
All XRD experiments were carried out at room temperature.
The pressure is cross checked with solid Ne diffraction peaks above \SI{5}{GPa} using the equation of state from Ref.~\cite{Dewaele2008}.
A gas membrane pressure controller was used to adjust the pressure.
In all the experiments, Ne was used as pressure medium.
The typical exposure time was 60 seconds per image.
The 2-D diffraction images were integrated using the DIOPTAS software~\cite{Prescher2015}.
LeBail refinements on the high pressure XRD data were performed in GSAS-II~\cite{Toby2013}. 

High-pressure dc-magnetization measurements were performed in a Quantum Design MPMS using a miniaturized Tozer-type turnbuckle DAC~\cite{Graf2011,Giriat2010}.
The diamonds had culets of \SI{1}{mm}. A Berylco25 gasket was pre-indented from 250 to \SI{100}{\micro\meter}.
The pressure medium was 1:1 n-pentane:isoamyl alcohol, which is known to be nearly hydrostatic to \SI{7.4}{GPa} at room temperature~\cite{Klotz2009}.
The ruby manometer signal was collected via fiber optic access through a custom sample rod.  Pressure was applied at room temperature for these measurements.

The ac-magnetic susceptibility measurements were performed using a balanced-primary/secondary-coil system that has been described elsewhere~\cite{deemyad_2001_1}.
The diamond anvil cell is an Almax-Easylab ``Chicago DAC''~\cite{Feng2014}, which is designed to fit inside the bore of a Quantum Design PPMS.
Samples with approximate dimensions of $\SI{200}{\um}\times\SI{200}{\um}\times\SI{50}{\um}$ were extracted from a larger chunk of polycrystalline material and loaded into the sample chamber. The diamonds had culets of \SI{800}{\micro\meter} and a Berylco25 gasket was indented from 250 to \SI{80}{\micro\meter}. Daphne oil was used as a pressure medium.
Two SR830 lock-in amplifiers are used in order to simultaneously measure the first and third harmonic of the ac magnetic susceptibility~\cite{Ishida1990}. The primary provides an excitation field of \SI{3}{Oe} RMS at \SI{1023}{\hertz}.
The detection coil is connected through a Stanford Research SR554 transformer/preamplifier.
For these measurements, the signal is dominated by a large, temperature-dependent background signal deriving from the gasket and nearby parts of the DAC.
In order to eliminate this contribution, we have measured the temperature-dependent background susceptibility and subtracted it from the data. 
The ruby manometer signal was collected via optical fiber and a lens system which is mounted to the diamond anvil cell.
Pressure changes were carried out at room temperature.

For the resistivity measurements, small pieces of sample with dimensions of about  $70\,\mathrm{\mu m} \times 70\,\mathrm{\mu m} \times 10\,\mathrm{\mu m}$ were cut from a larger piece of polycrystal for each of the measurements.
The  measurements were carried out in a OmniDAC gas-membrane-driven diamond anvil cell from Almax-EasyLab. 
The cell was placed inside a custom, continuous-flow cryostat built by Oxford Instruments.
Optical access to the cell for visual observation and measurement of the ruby manometer is provided through windows at the bottom of the cryostat and an optical fiber entering via a feed-through at the top.
One of the diamonds used was a designer-diamond anvil containing eight symmetrically arranged, deposited-tungsten microprobes encapsulated in high-quality-homoepitaxial diamond~\cite{weir_2000_1}.
This diamond had a culet diameter of $\sim 180\,\mathrm{\mu m}$, and the opposing anvil had a culet diameter of $\sim 500\,\mathrm{\mu m}$. 
Resistance was measured in the van der Pauw geometry with currents of $\sim$1 mA. Gaskets were pre-indented from $150\,\mathrm{\mu m}$ to $\sim 30\,\mathrm{\mu m}$ thickness and were made of 316 stainless steel. 
Quasihydrostic soft, solid steatite was used as the pressure-transmitting medium.
The temperature at which pressure was applied varied in different experimental runs as specified in the text and figures.

\section{Crystal structure}
\begin{figure}
	\includegraphics[width=\columnwidth]{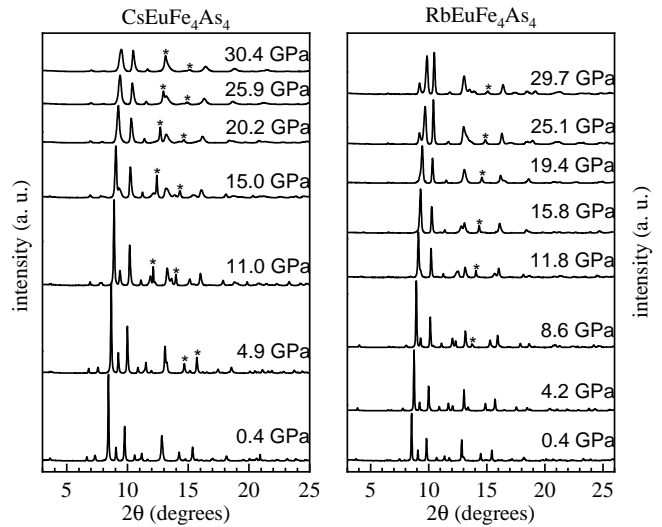}
    \caption{Selected XRD patterns for \Cs\ (left) and \Rb\ (right) at various pressures. The data were taken at room temperature. Solid Ne peaks are identified by asterisks in the spectra at and above \SI{4.9}{GPa}.}
    \label{fig:xrd}
\end{figure}

\begin{figure}
	\includegraphics[width=\columnwidth]{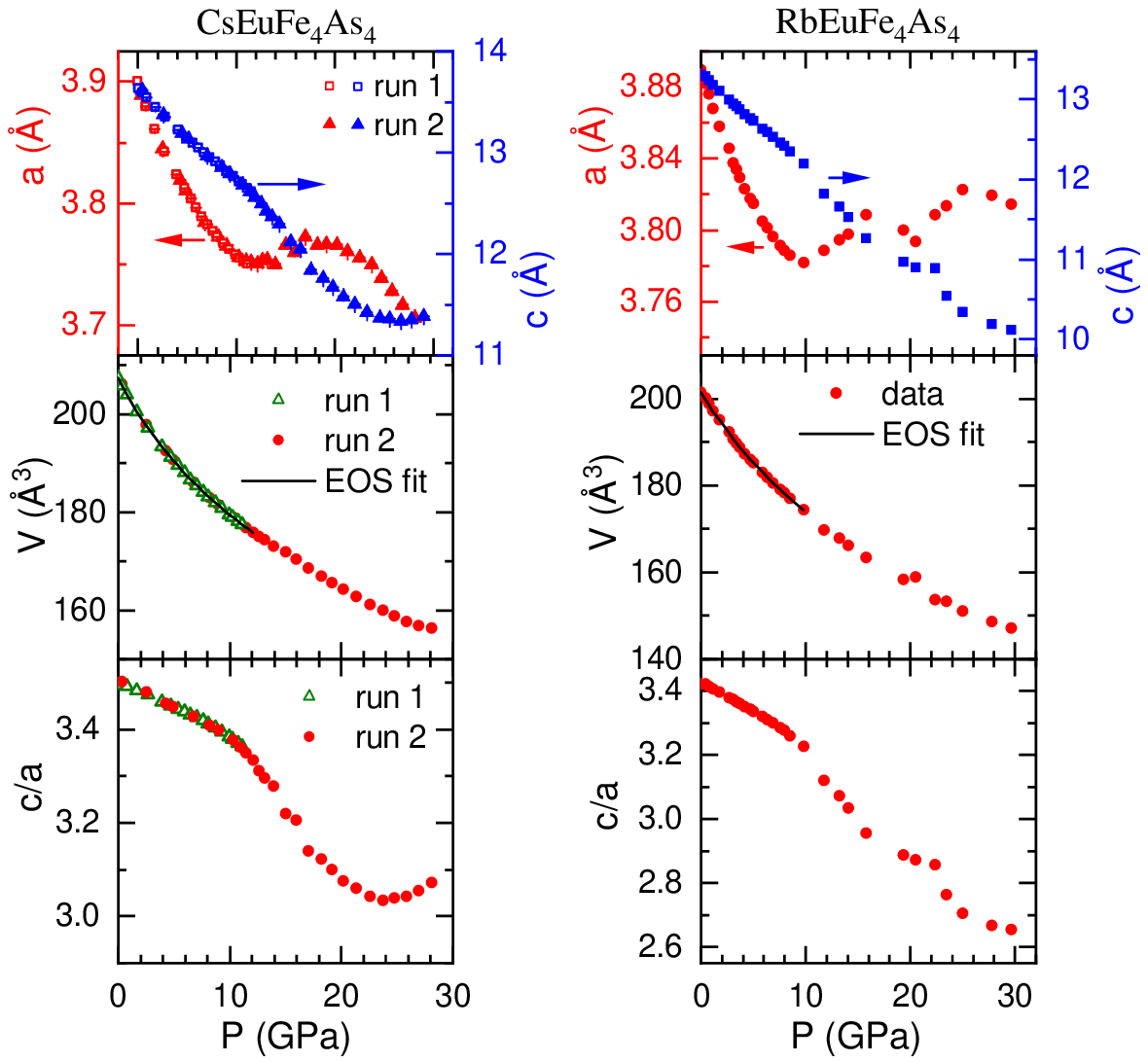}
    \caption{Lattice parameters, unit cell volume, and $c/a$ ratio versus pressure for \Cs\ (left) and \Rb\ (right).}
    \label{fig:eos}
\end{figure}

Representative XRD patterns are shown in Fig.~\ref{fig:xrd}.
At $\sim \SI{5}{GPa}$ additional peaks from solid Ne pressure medium (marked by asterisks) appear in the diffraction pattern.
The lattice parameters $a$ and $c$ obtained from LeBail refinements with space group P4/mmm are shown in Fig.~\ref{fig:eos}.
Both $a$ and $c$ decrease smoothly with pressure up to \SI{12}{GPa} (\Cs) and \SI{10}{GPa} (\Rb).
The volume as a function of pressure in the low pressure structure for both compounds is fit with the third-order Birch-Murnaghan equation~\cite{Birch1978}.
For \Cs, we find $B_0 = \SI{46.3(2)}{GPa}$ and $B_0^{\prime} = 5.59(6)$.
For \Rb, we find $B_0 = \SI{50.7(7)}{GPa}$ and $B_0^{\prime} = 4.1(3)$.
Due to the absorption of the c-BN seat and spotty nature of the data, Rietveld refinement was not successful.

With further increase of pressure, anomalous compression is evidenced by the negative compressibility of $a$ and a gradual collapse of the $c$ lattice constant evidenced by a change in the slope of $c$ vs $P$.
These features are also visible  in the $c/a$ ratio plotted in the lower panels of Fig.~\ref{fig:eos}.
Such tetragonal to collapsed-tetragonal structural transitions are common in 122 compounds~\cite{Kreyssig2008,Jørgensen2009,Uhoya2010,Jayasekara2015}.
Similar anomalous compression has been observed in CaKFe$_4$As$_4$ near \SI{4}{GPa}~\cite{Kaluarachchi2017} and in the 122 analog EuFe$_2$As$_2$ around $8-\SI{12}{GPa}$~\cite{Uhoya2010,Yu2014}.
The relatively small changes in the lattice parameters at the structural transition for \CaK, combined with density functional theory (DFT) calculations, lead the authors of Ref.~\cite{Kaluarachchi2017} to propose that the transition is to a ``half-collapsed'' tetragonal phase.
In this phase, As-As bonds form across the Ca layer, but not across the K layer.
At still higher pressures, one might thus expect another collapse transition as As-As bonds form across the K layer.
In the present results on \Rb\ and \Cs\ we find changes in the $a$ and $c$ lattice constants that are consistent with the half collapse scenario since they are comparable to those found for \CaK\ and substantially smaller than those found for KFe$_2$As$_2$~\cite{Nakajima2015} and CaFe$_2$As$_2$~\cite{Kreyssig2008}.

In the case of the \Rb\ data, we see two anomalies in the lattice constant vs pressure data.
This is most clearly visible in the $a$ vs P data, where $a$ begins to increase at $\sim \SI{10}{GPa}$, passes through a maximum and then again begins to increase with pressure near \SI{20}{GPa}.
These two features may be connected with a sequence of transitions from tetragonal (T) to half-collapsed (hcT) starting at \SI{10}{GPa}, followed by a transition from hcT to fully collapsed tetragonal (cT) starting at \SI{20}{GPa}.
In the case of \Cs, the data suggest that the transition to the hcT phase begins at \SI{12}{GPa}, while an eventual cT phase likely appears at a pressure somewhere above \SI{30}{GPa}.

\section{Magnetic measurements}
\begin{figure}	
	\includegraphics[width=\columnwidth]{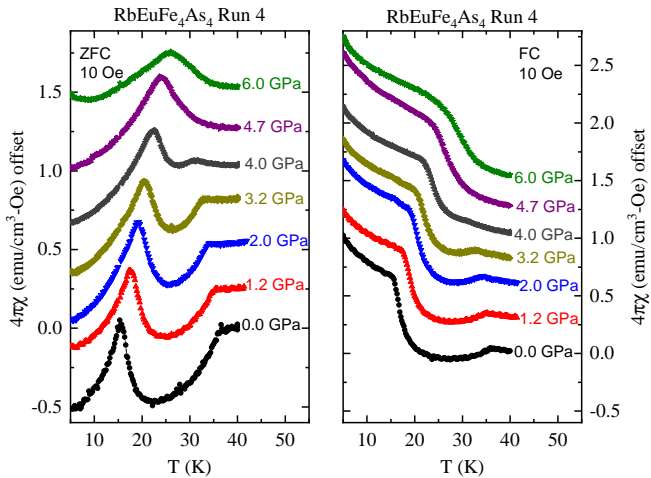}
	\caption{Measurements of the dc susceptibility of \Rb\ plotted vs temperature for several different pressures. The left and right show the zero field cooled (ZFC) and field cooled (FC) data. The data have been offset for clarity.}
	\label{fig:DC}
\end{figure}
Figure~\ref{fig:DC} presents the results of dc magnetization measurements on \Rb\ at several values of the applied pressure.
The data were collected with an applied field of \SI{10}{Oe}.  
The zero-field-cooled (ZFC) and field cooled (FC) measurements are plotted in separate panels and the data have been offset for clarity.
The measurement marked \SI{0.0}{GPa} was collected with the sample loaded inside the pressure cell sample chamber.
At zero pressure, both the superconducting transition ($T_c \sim \SI{36}{K}$) and the magnetic transition ($T_m \sim \SI{15}{K}$) are clearly visible.
The data have been plotted using the estimated volume of the sample such that a full Meissner effect would generate a change in the signal of $-1$.
The measurements indicate a shielding fraction of $\sim 45\%$, which is consistent with the relatively large size of the ferromagnetic anomaly compared to the superconducting drop.
We note that the exact volume of the tiny, irregularly shaped sample is difficult to estimate precisely and the error in this calibration could be as large as $\sim 50\%$.
The type-II nature of the superconductivity is evident from the smaller magnitude of the diamagnetic drop in $\chi$ for the FC measurements compared to the ZFC measurements.
\begin{figure}
	\includegraphics[width=\columnwidth]{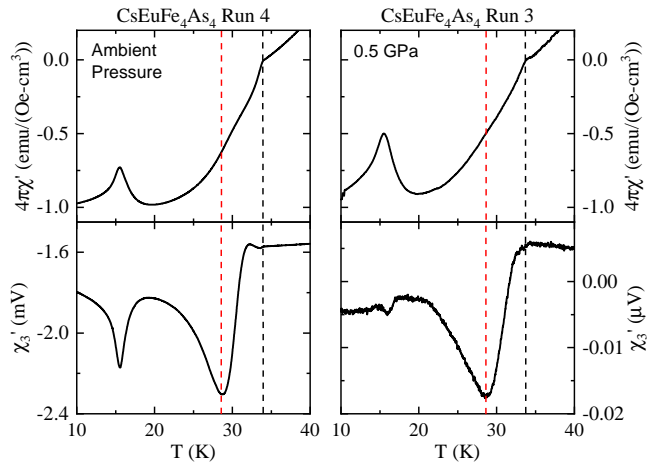}
    \caption{Real parts of the first and third harmonic as a function of temperature for a large piece of \Cs\ at ambient pressure (left) and a small piece inside the diamond anvil cell (right). The vertical dashed lines are guides to the eye showing the correspondence between features in the first and third harmonic.  The minimum in $\chi_3'$ occurs near the midpoint of the superconducting transition in \acfirst.}
    \label{fig:3rd}
\end{figure}

As pressure increases, the magnetic transition moves to higher temperature and the diamagnetic signal at \Tc\ moves to lower temperature and becomes smaller.
Somewhere between 4.0 and $\SI{4.7}{GPa}$ clear signatures of the superconducting transition vanish.
Extrapolation of the two transition temperatures suggests that they do not meet until $\sim \SI{7}{GPa}$.
The disappearance of the superconducting signal at a somewhat lower pressure might be due to substantial flux pinning on cooling through the superconducting transition when \Tm\ is just below \Tc.

Figure \ref{fig:3rd} shows a comparison of the real parts of the first (\acfirst) and third harmonic (\acthird) of the ac susceptibility for \Cs.
Results are shown for both a large piece of sample at ambient pressure and for a small piece of sample loaded in the diamond cell at the lowest applied pressure, \SI{0.5}{GPa}.
The data have been plotted in units such that $4\pi\chi = -1$ corresponds to full shielding, by using estimates of the sample volume.
Both the ambient pressure sample and the sample loaded in the diamond cell show diamagnetic signals that are consistent with full shielding.

While the interpretation of \acfirst\ is simple, the interpretation of \acthird\ is less straightforward.
It is known that the shape of the \acthird\ transition can have a complicated dependence on measurement conditions.
Analysis of the frequency dependence of \acthird\ can provide insight into the flux dynamics of the material~\cite{Gioacchino2010,DiGioacchino1999}. 
From a practical standpoint, \acthird\ provides a complementary way to track the transition temperatures as a function of pressure.
Figure~\ref{fig:3rd} demonstrates that the superconducting onset temperature in \acfirst\ occurs at approximately the same temperature as the onset in \acthird.
In addition, the minimum in \acthird\ just below \SI{30}{K} is in good agreement with the midpoint of the superconducting transition measured via \acfirst.
As we will see later, the minimum in \acthird\ also corresponds very closely with the midpoint of the resistive transition.
Though the feature at \Tm\ is visible in the high pressure data, it is substantially smaller in relative magnitude than the corresponding feature in the large, ambient pressure sample.
\begin{figure}	
	\includegraphics[width=\columnwidth]{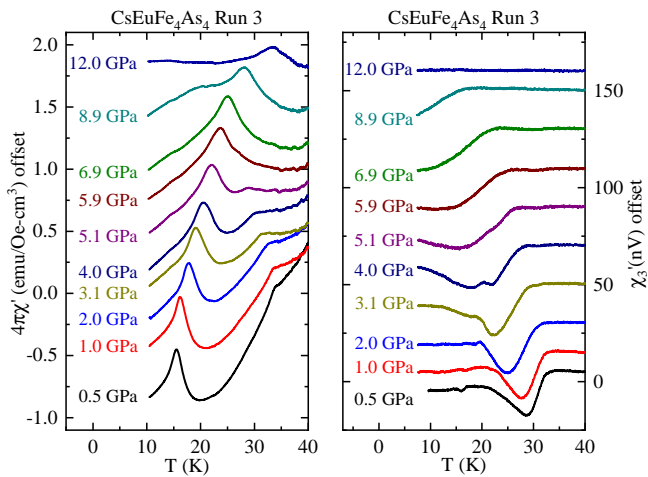}
	\caption{AC magnetic susceptibility as a function of temperature for \Cs. (a) Real part of the first harmonic (\acfirst). The superconducting transition is visible as a drop in the signal and the magnetic transition corresponds to the peak. (b) Real part of the third harmonic (\acthird).}
    \label{fig:AC}
\end{figure}

Having established the approximate shielding fraction and relationship between first and third harmonic signals, we now turn to the high-pressure ac susceptibility results.
Figure~\ref{fig:AC} shows a selection of ac magnetic susceptibility measurements for \Cs.
The trends are very similar to those observed in the dc measurements on \Rb, though the measurements extend to higher pressures.
Increasing pressure causes an increase in \Tm\ and a decrease in \Tc.
When the superconducting and magnetic transition are very close in temperature (\SI{5.9}{GPa} and \SI{6.9}{GPa}) it becomes difficult to distinguish the location of \Tc.
However at higher pressure (\SI{8.9}{GPa}) there appears to be a diamagnetic superconducting signal just below \SI{20}{K}, which is lower than $T_m \sim \SI{28}{K}$.
This suggests that a significant fraction of the sample remains superconducting when $T_m > T_c$.
However, the proximity of \Tm\ and $T_c$, together with the broadness of the superconducting transitions, make it impossible to obtain an unambiguous estimate of the superconducting volume fraction at the higher pressures.
Nonetheless, by \SI{12}{GPa}, which is above the structural transition, there is clearly no trace of a superconducting transition - though the signal at \Tm\ remains.

Figure \ref{fig:AC}b shows the corresponding measurements of \acthird, which were measured simultaneously with the first harmonic at each pressure.
An onset in the \acthird\ signal is still visible up to \SI{8.9}{GPa}, but is supressed to below \SI{5}{K} by \SI{12}{GPa}.
The minimum in \acthird, which corresponds approximately to the midpoint of the superconducting transition, can be tracked to \SI{5.9}{GPa}.
Both onset and midpoint indicate a monotonic suppression of \Tc\ with pressure.

\section{Electrical resistivity measurements}
\begin{figure}
	\includegraphics[width=\columnwidth]{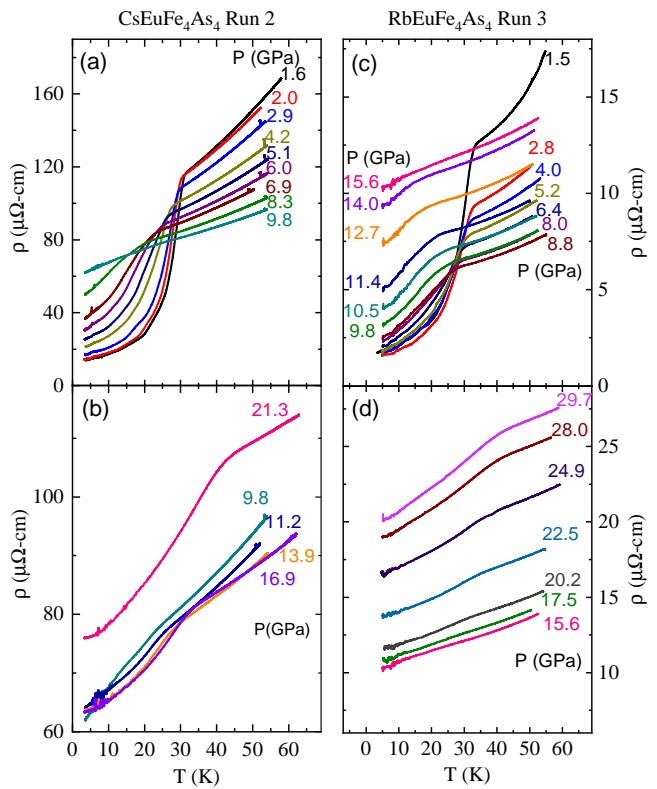}
    \caption{Resistivity as a function of temperature at various pressures for \Cs\ (panels a and b) and \Rb\ (panels c and d).  For clarity, the curves corresponding to superconducting transitions are plotted in the top panels and those corresponding to magnetic transitions are plotted in the bottom panels.  The superconducting transition is suppressed by pressure, while the magnetic transition is enhanced.}
    \label{fig:rhoSC}
\end{figure}
Figure~\ref{fig:rhoSC} shows resistivity versus temperature curves for \Cs\ and \Rb\ at selected pressures spanning the entire pressure range studied.
In order to present the data clearly and avoid excessive overlapping of the curves, we have plotted the data corresponding to superconducting transitions in the upper panel, and those corresponding to magnetic transitions in the lower panel.
The resistivity does not drop completely to zero for either compound.
At the lowest temperatures, roughly 20\% of the normal state resistance remains.
We tested 3-4 samples of each compound at the lowest achievable pressure ($\sim 1-\SI{2}{GPa}$) and never achieved zero resistance.
Typically, at the lowest temperatures, the resistance dropped to 20-40\% of the normal state resistance above \Tc\ (though in one case, the drop was only 10\%).
The absence of zero resistance can not be an artifact of the measurement technique, since similar measurements on superconductors with the same designer anvil have produced zero resistance~\cite{VanGennep2017,Jackson2017}.
It is possible that the substantial shear forces associated with the solid-steatite pressure medium contribute to a poor connectivity and sizable inter-grain resistance even below \Tc.
We also note that the low pressure values of the resistivity vary substantially from sample to sample.
This is likely related to several factors: the uncertain geometry of the very tiny samples, the varying strain in the solid pressure medium, and possible impurity phases in the polycrystalline samples.
High pressure measurements on single crystals would likely resolve these issues.

\begin{figure}
    \includegraphics[width=\columnwidth]{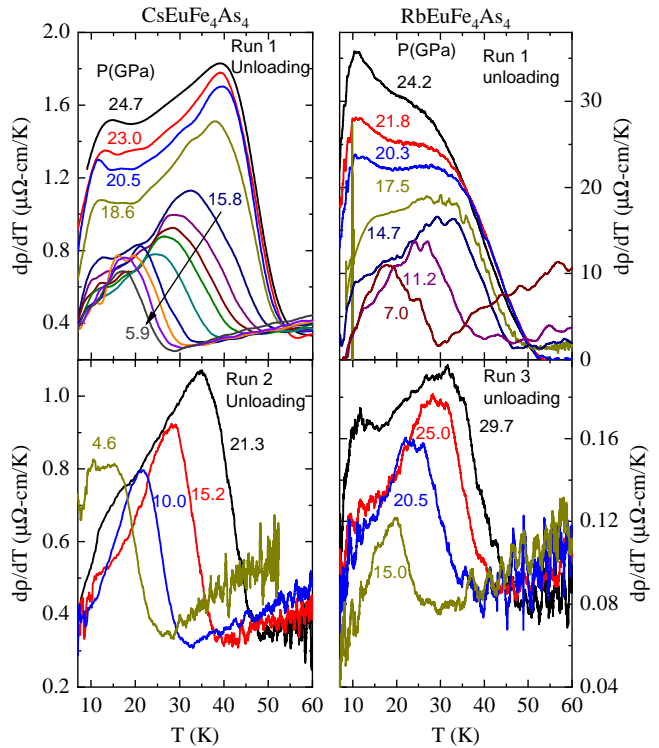}
    \caption{Derivative of the resistivity as a function of temperature for several different pressures while unloading in the regime where $T_m > T_c$. Data for two different pressure runs are shown for both \Cs\ (left) and \Rb\ (right).}
 	\label{fig:rhoFM}
\end{figure}

Despite the issues described above, the resistivity data show clear trends that allow us to track both \Tc\ and \Tm\ as a function of pressure.
For both compounds \Tc\ decreases smoothly with pressure.
The drop in the resistance above \Tc\ becomes smaller at higher pressures and eventually vanishes.
Once the signatures of superconductivity vanish, another weaker anomaly appears in the resistivity.
This feature moves to higher temperature and becomes more pronounced with increasing pressure.
At lower pressures in particular, the weaker anomaly is difficult to see in the raw resistivity data (Fig.~\ref{fig:rhoSC}b,d), but is clearly visible in the derivative of this data.
Derivative ($d\rho/dT$) data are presented for several different experimental runs in Fig.~\ref{fig:rhoFM}.
The high pressure anomaly in the resistivity is clearly due to the magnetic ordering transition, since it shows the same pressure dependence as \Tm\ (see Fig.~\ref{fig:phase}).

\section{Phase diagrams}
\begin{figure*}
	\includegraphics[width=\columnwidth]{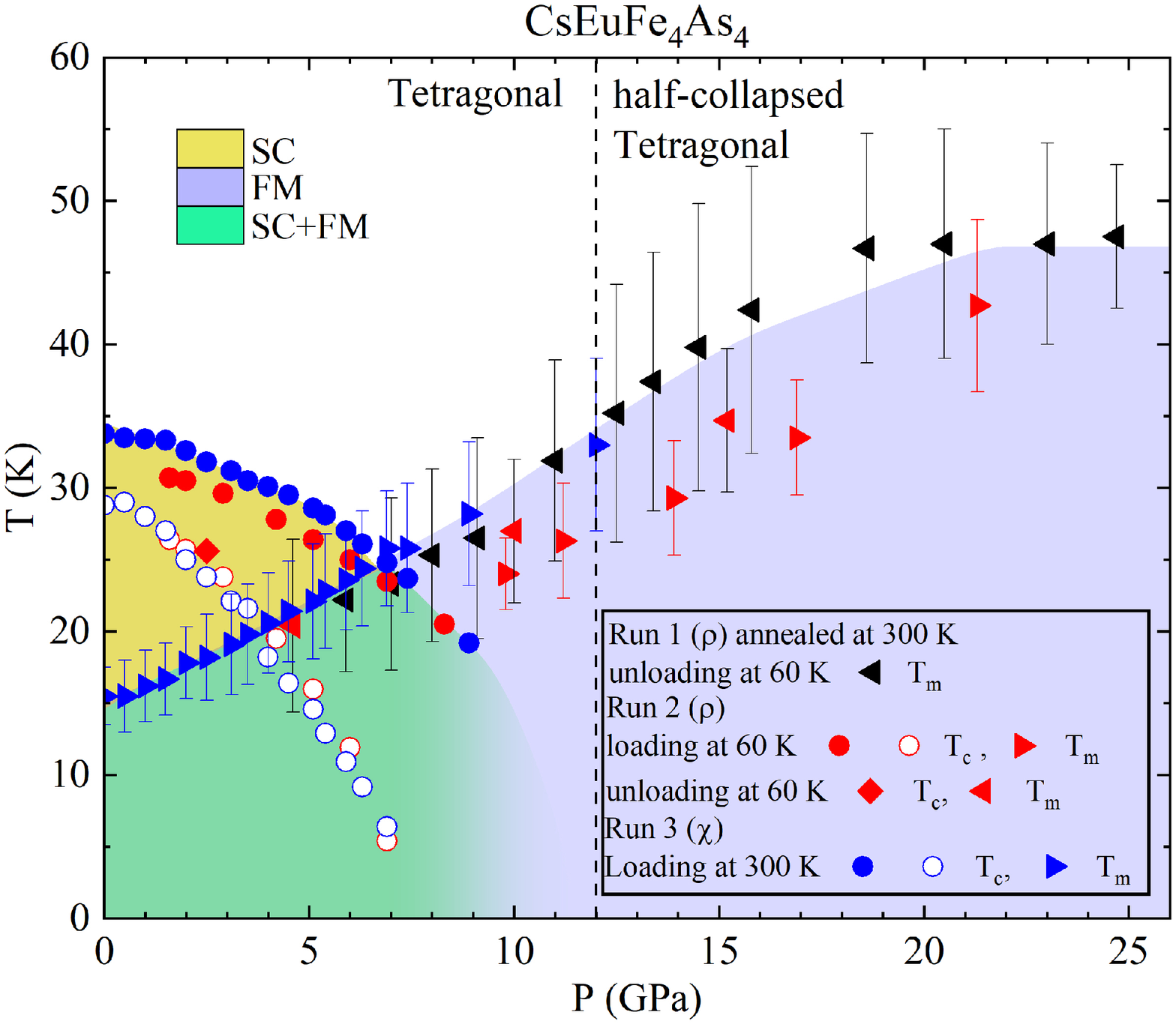}
    \hspace{0.05in}
    \includegraphics[width=\columnwidth]{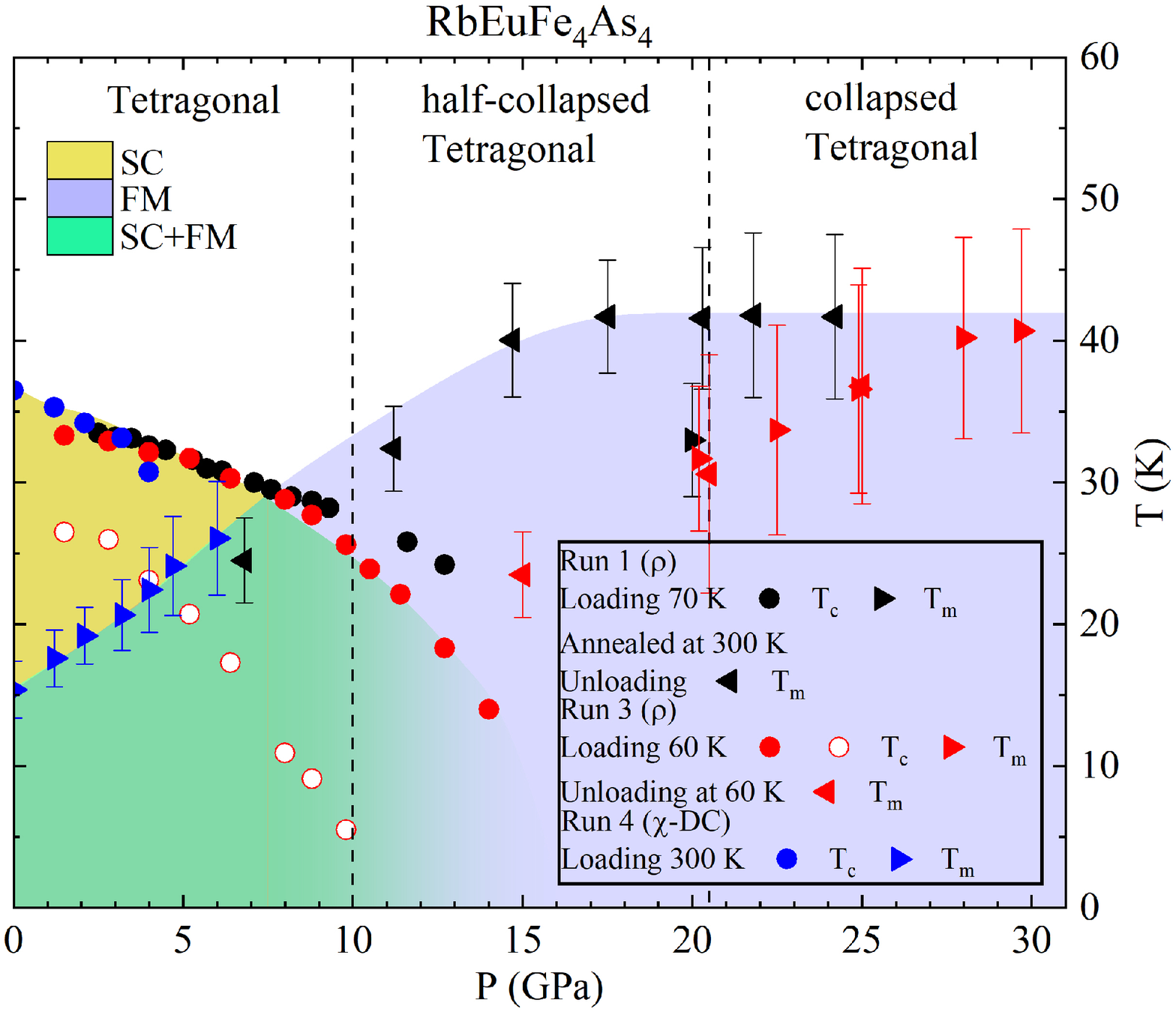}
    \caption{Superconducting and magnetic phase diagram of \Cs\ (left) and \Rb\ (right). The open symbols correspond to the midpoints of the superconducting transition, as described in the text. Superconducting transition temperatures (solid circles) correspond to the onset of the transition measured via resistivity or magnetic susceptibility.  Magnetic transition temperatures (solid triangles) are determined from the peak in magnetic susceptibility or midpoint of the feature in $d\rho/dT$. The dashed, vertical lines indicate onset pressures for the structural transitions.}
    \label{fig:phase}
\end{figure*}
Combining the data from six different pressure runs, consisting of magnetic or resistivity measurements at more than 90 different pressures, together with the room-temperature x-ray diffraction data, we arrive at the phase diagrams presented in Figure~\ref{fig:phase}.
The dashed vertical lines indicate the critical pressures for the onset of the structural transitions, which have been determined at room temperature from the pressure at which the $a$ lattice constant begins to increase.
While it is possible that these transition pressures have some temperature dependence, in the case of \CaK, it was found that the T$\rightarrow$hcT transition pressure was not strongly dependent on temperature~\cite{Kaluarachchi2017}.
Both \Cs\ and \Rb\ exhibit quite similar phase diagrams.
The similarity in the phase diagrams for both compounds is not surprising given that, at $P=0$, the values of $T_m$, \Tc, and the compressibility of the two compounds are nearly identical.

Under pressure, \Tm\ increases to a maximum value near $\sim 40-\SI{50}{K}$ at $\sim 25-\SI{30}{GPa}$.
The initial ($P \sim 0$) values of the slope $dT_m/dP$ are $\sim \SI{1.7}{K/GPa}$ for \Rb\ and $\sim \SI{1.5}{K/GPa}$ for \Cs.
We do not detect any significant anomaly or change in slope of the \Tm\ vs pressure curve at the onset of the T$\rightarrow$hcT transition.
With increasing pressure, \Tc\ is suppressed monotonically.
Assembling data from multiple measurements suggests that the onset of the superconducting transition falls below \Tm\ at $P_c \sim \SI{7}{GPa}$ for both compounds.
In the case of the Cs compound, we have evidence from a single temperature sweep that a substantial fraction of the sample remains superconducting when $T_c > T_m$ (see \acfirst\ and \acthird\ data for \SI{8.9}{GPa} in Fig.~\ref{fig:AC}).

The criteria for the superconducting onset temperature is given by the intersection of linear fits to the data just above and just below the the onset of the transition.
Magnetic ordering temperatures, \Tm, are determined using the peak in $\chi$ (Figs.~\ref{fig:DC} and \ref{fig:AC}), or the the midpoint of the feature in $d\rho/dT$ (Fig.~\ref{fig:rhoFM}).
The open symbols in the phase diagrams correspond to the midpoint of the superconducting transition.
The midpoint values are estimated by taking either the temperature where the resistivity has dropped to 50\% of the normal state value just above the onset or by taking the minimum in \acthird\ (see Fig.~\ref{fig:3rd}).
For the \Cs\ phase diagram, where both types of data are available, there is very good agreement between the superconducting transition midpoints determined from resistivity and \acthird.
In the case of \Cs, the midpoint goes to zero roughly at $P_c$.
For \Rb, the midpoint of the superconducting transition vanishes at a slightly higher pressure ($\sim \SI{10}{GPa}$).

For some of the measurements, the sample was annealed at room temperature at high pressure, while for other measurements pressure application occurred at low temperature and the sample was maintained at $\sim 60-\SI{70}{K}$ throughout the course of the experiment (see key to Fig.~\ref{fig:phase}).
For the superconducting transition, the data are in good agreement for both types of pressure application.
The data may suggest that samples compressed at low temperatures tend to present the magnetic transition at a lower temperature than samples subjected to high pressure at room temperature.
This can be seen by comparing the black and red triangles in the high pressure part of the phase diagram.
The effect is most pronounced for the \Rb\ sample, though a small effect with the same trend appears to exist in the \Cs\ data as well.
One possibility is that \Tm\ is sensitive to the hydrostaticity of the pressure conditions.
Annealing the sample at room temperature under pressure may tend to relieve strain in the sample.
The phase diagrams of several 122-type iron-based superconductors are well known to be sensitive to the degree of hydrostaticity~\cite{Paglione2010,Stewart2011}.
Another possibility is that kinetic effects due to low temperature compression alter the pressures ranges for which different crystal structures are present in the sample~\cite{Jackson2017,Desgreniers2015}.
There are not yet low-temperature/high-pressure x-ray diffraction studies on $A$Eu\FeAs\ compounds to test whether this might be the case.

\section{Discussion}
Among 1144 materials, to date, only CaKFe$_4$As$_4$ appears to have been the subject of a study under applied pressure~\cite{Kaluarachchi2017}.
For CaKFe$_4$As$_4$, pressure causes a similar decrease in critical temperature with pressure, though at a slightly higher rate than we find for \Rb\ and \Cs.
At \SI{4}{GPa} the structure collapses to an hcT phase, and bulk superconductivity vanishes (though traces of filamentary superconductivity remain)~\cite{Kaluarachchi2017}.
For (Rb,Cs)Eu\FeAs, an important question is whether bulk superconductivity begins to vanish at the onset of the structural transition or, perhaps, at \Pc, where \Tc\ dips below \Tm.
The former possibility seems probable since we see substantial signatures of superconductivity in the susceptibility for \Cs\ at pressures above \Pc.

Another interesting question is whether the initial T$\rightarrow$hcT transition corresponds to As-As bonds developing across the Eu layer or, alternatively, across the Rb/Cs layer.
Analysis of our x-ray data has not allowed us to unambiguously choose between these possibilities.
Comparison with the behavior of other 122 compounds at high pressure~\cite{Huhnt1998,Terashima2009,Matsubayashi2011,Kurita2011b,Kurita2012,Kobayashi2012,Kumar2014} gives some insight.
Several Eu-based 122 compounds are known to exhibit pressure-induced T$\rightarrow$cT transitions that are connected to a valence change from Eu$^{2+}$ to non-magnetic Eu$^{3+}$~\cite{Huhnt1998,Ni2001,Uhoya2010}.
For comparison with \Cs\ and \Rb, we first look to \Eu.

At ambient pressure, \Eu\ exhibits antiferromagnetic order (\TN$\sim \SI{20}{K}$) deriving from Eu$^{2+}$ ions.
\Eu\ exhibits pressure induced superconductivity (coexisting with antiferromagnetic order) over a narrow range of pressures near \SI{2}{GPa}~\cite{Miclea2009}.
Under pressure \TN\ eventually begins to increase and then transforms to ferromagnetism at $\sim \SI{6}{GPa}$~\cite{Matsubayashi2011}.
A T$\rightarrow$cT transition commences at $\sim \SI{10}{GPa}$ (at low temperature)~\cite{Yu2014}.
At roughly the same pressure, the ferromagnetic ordering temperature passes through a maximum and begins to decrease.
The magnetic order and moment of the Eu ion appear to be completely suppressed by \SI{20}{GPa}~\cite{Matsubayashi2011}.
These observations are consistent with a valence transition from Eu$^{2+}$ to non-magnetic Eu$^{3+}$ that commences near the structural transition, but is not complete until significantly higher pressures.
Examination of the phase diagrams in Fig~\ref{fig:phase}, shows that the pressure dependence of the magnetic ordering temperature does not exhibit any clear anomaly at the onset of the T$\rightarrow$hcT transitions.
On the other hand, it does appear that \Tm\ begins to saturate within the hcT phases.
Therefore, from our \Tm\ versus $P$ data alone, it is not possible to categorically select which layer (Eu or alkali metal) initially collapses.
However, as discussed below, consideration of structural trends in iron-based 122 compounds strongly suggests that the initial collapse occurs in the Eu layer.

Yu \textit{et al.}~\cite{Yu2014} examined the critical pressure for the T$\rightarrow$cT transition pressure in AFe$_2$As$_2$ compounds ($A$ = Ca, Sr, Ba, and Eu).
They noted that the critical pressure showed a trend of increasing with increasing cation radius.
This trend is consistent with the findings of DFT calculations on \CaK~\cite{Kaluarachchi2017}, which indicate that the Ca layer collapses first ($r_{Ca^{2+}} = \SI{1.0}{\angstrom}$) while the K layer only collapses at higher pressures ($r_{K^{+}} = \SI{1.4}{\angstrom}$)~\cite{SpringerHandbook}.
The ionic radius of Eu$^{2+}$ (\SI{1.2}{\angstrom}) is smaller than that of both Rb$^{+}$ (\SI{1.5}{\angstrom}) and Cs$^{+}$ (\SI{1.7}{\angstrom})~\cite{SpringerHandbook}.
Consequently, in \Cs\ and \Rb, the collapse of the Eu layer should occur first, with the alkali metal layer collapsing at higher pressure.
This picture is also consistent with our observation that the sample containing the smaller Rb$^+$ ion exhibits a second collapse transition beginning at $\sim \SI{20}{GPa}$, while for the sample containing the larger Cs$^+$ ion, the second collapse does not occur below \SI{30}{GPa}.
Comparison with the trend of the T$\rightarrow$cT pressure versus ionic radius presented in Ref.~\cite{Yu2014} suggests that the Cs layer in \Cs\ should collapse at a pressure of $\sim \SI{30}{GPa}$, which is just at the limit of the range of our measurements.

\section{Conclusion} 
In summary, we find that both \Rb\ and \Cs\ exhibit very similar phase diagrams under pressure.
X-ray diffraction measurements suggest a transition to a half-collapsed tetragonal phase at pressures of \SI{10}{GPa} and \SI{12}{GPa} for \Rb\ and \Cs, respectively.
For \Rb, an additional structural transition to a fully collapsed tetragonal phase may occur at \SI{20}{GPa}.
For both materials, the magnetic transition temperature, $T_m$, increases with pressure while the superconducting transition temperature \Tc\ decreases.
The two transitions coincide near a critical pressure $P_c \sim \SI{7}{GPa}$, indicating that a crossover from FMS to SFM occurs prior to the onset of the tetragonal $\rightarrow$ half-collapsed-tetragonal transition.
The relatively modest pressures required to tune this crossover make $A$Eu\FeAs\ compounds a very interesting system to further explore the interplay of superconductivity and (local moment) magnetism in a clean, stoichiometric material.

The present measurements have been performed using polycrystalline samples which show somewhat broad transitions.
Recently, single crystalline specimens~\cite{Bao2018} have been grown which show substantially sharper transitions.
It would be particularly interesting to further examine the narrow pressure range around $P_c$ in such crystals in order to explore \textit{e.g.}, the influence of the FMS to SFM crossover on the upper critical field curve.

\section*{Acknowledgments}
This work was supported by National Science Foundation (NSF) CAREER award DMR-1453752.  High pressure technique development was partially supported by a National High Magnetic Field Laboratory User Collaboration Grant.  The National High Magnetic Field Laboratory is supported by the NSF via Cooperative agreement No.\ DMR-1157490, the State of Florida, and the U.S. Department of Energy. Designer diamond anvils were supported by DOE-NNSA Grant No.\ DE-NA0002928 and under the auspices of the U.S. Department of Energy by Lawrence Livermore National Laboratory under Contract DE-AC52-07NA27344. Support from COMPRES under NSF Cooperative Agreement EAR-1606856 is acknowledged for COMPRES-GSECARS gas loading system, the PX2 program and partial support of W.\ Bi.\ P.\ Materne would like to acknowledge support from Deutsche Forschungsgemeinschaft (DFG, German Research Foundation) in the project MA 7362/1-1. We thank S.\ Tkachev for help with the Ne loading of the DACs at the APS. Y.\ Liu and G.\ Cao acknowledge the support by the National Science Foundation of China (under Grants No.\ 11474252). We thank G.\ R.\ Stewart, K.\ Quader, P.\ J.\ Hirschfeld, and R.\ Valent\'i for helpful comments.

\bibliography{AEuFe4As4}
\end{document}